\documentclass[twocolumn]{aa}
\newcommand{\mr}{\mathrm}

\usepackage{graphicx,txfonts,natbib,verbatim}
\bibpunct{(}{)}{;}{a}{}{,}

\begin{document}

\title{Wave-particle interactions in non-uniform plasma and the interpretation of Hard X-ray spectra in solar flares}

\author{E. P. Kontar, H. Ratcliffe, and N.H. Bian}

\offprints{E.P. Kontar \email{Eduard.Kontar@glasgow.ac.uk}}

\institute{School of Physics \& Astronomy, University of Glasgow, G12 8QQ, United Kingdom}

\date{Received ; Accepted }

\abstract{High energy electrons accelerated during solar flare are abundant in the solar corona and in the interplanetary space.
Commonly, the number and the energy of non-thermal electrons at the Sun is estimated using hard X-ray (HXR)
spectral observations (e.g. RHESSI) and a single-particle collisional approximation. }
{To investigate the role of the spectrally evolving Langmuir turbulence on the population of energetic electrons
in the solar corona.}
{We numerically simulate the relaxation of a power-law non-thermal electron population in a collisional inhomogeneous plasma
including wave-particle, and wave-wave interactions.}
{The numerical simulations show that the long-time evolution of electron population above 20 keV
deviates substantially from the collisional approximation when wave-particle interactions in non-uniform plasma are taken into account.
The evolution of Langmuir wave spectrum towards smaller wavenumbers, due to large-scale density fluctuations
and wave-wave interactions, leads to an effective acceleration of electrons. Furthermore, the time-integrated spectrum
of non-thermal electrons, which is normally observed with HXR above 20 keV, is noticeably increased
due to acceleration of non-thermal electrons by Langmuir waves.}
{The results show that the observed HXR spectrum, when interpreted in terms of collisional relaxation,
can lead to an overestimated number and energy of energetic electrons accelerated in the corona.}


\titlerunning{Inhomogeneous plasma evolution of energetic electrons}

\authorrunning{Kontar et al}

\maketitle

\section{Introduction}

Solar flares provide a number of theoretical challenges for various aspects of electron
dynamics, including their acceleration and transport. One of the important aspect of solar flares
is the high acceleration efficiency of electrons, seen via their Hard X-ray emission (HXR)
in the dense atmosphere footpoints \citep[e.g.][]{1990ApJS...73..343B,1999Ap&SS.264..129S,2002SSRv..101....1A,2003AdSpR..32.1001L,2006SSRv..124..233L,2010A&A...519A.114B,2011A&A...526A...3B,2011A&A...533L...2B,2011SoPh..270..493C,2011ApJ...739...96K}.
It was realized early on that the solar corona can store large amounts of
magnetic energy \citep{1958IAUS....6..123S,2002A&ARv..10..313P},
that can be released to increase the kinetic
energy of the surrounding plasma.  While the exact location and properties of the electron
acceleration site are still under debate, spatially resolved HXR observations,
notably with RHESSI, point to the coronal origin of electron acceleration
and subsequent transport of energetic electrons towards the denser lower
corona \citep[e.g.][as the recent reviews]{2011SSRv..159..107H,2011SSRv..159..301K}.
HXR emission from the dense regions is then used as the model dependent
diagnostic of electron acceleration in the corona. The low dynamic range
of current hard X-ray instruments is insufficient to observe the evolution
of the energetic electron distribution {\it enroute} between the coronal and the footpoint parts.
The exception to this rule is probably the radio observations of emissions
from energetic electrons \citep[e.g.][]{2011ApJ...731L..19F}.

The propagation of energetic electrons between the coronal and the footpoint
sources is often treated as a simple collisional transport.
In this approximation the energy of an electron $E=m{\mr v}^2/2$ decreases with time
due to binary Coulomb collisions,
so that for non-thermal particles ${\mr v}(t) \gg{\mr v}_{Te}$, one
can write
\begin{equation}\label{dedx_TT}
\frac{d{\mr v}}{dt}=-\frac{\Gamma} {{\mr v}^{2}},
\end{equation}
where $\Gamma=4\pi e^4 n\ln \Lambda/m^2$, $n$ is the number density of the surrounding plasma,
$e$ is the electron charge, $m$ is electron mass, ${\mr v}_{Te}$ is the electron thermal
velocity, and $\ln \Lambda \simeq20$ is the Coulomb logarithm. This treatment obviously ignores all collective
interactions in plasma, but gives a simple relation between the accelerated
and the resulting mean electron spectra responsible for the observed HXR
spectra \citep[e.g.][]{2003ApJ...595L.115B}\footnote{Normally, only spatial dependency
$x={\mr v}t$ instead of temporal $t$ is considered.}.

Realizing the importance of various streaming instabilities and the subsequent
wave-particle interactions, a number of  models have included the generation of Langmuir waves.
Indeed, Langmuir wave turbulence can be effectively generated
when the bump on tail distribution is formed \citep{1973plas.book.....K,1995lnlp.book.....T}.
Firstly, the initially stable $\partial f/\partial {\mr v} < 0$ electron distribution function $f({\mr v})$ can
be turned unstable i.e. $\partial f/\partial {\mr v} >0$, when the fast electrons overtake the slower
ones. In the case of collisional relaxation of the electron population,
since slower particles lose their energy faster, $\tau _{col}(E)\simeq m{\mr v}^3/(2\Gamma)$,
a ``gap'' distribution  can appear from an initially power-law distribution on collisional time-scales.
This ``gap'' distribution of electrons creates a positive slope $\partial f/\partial{\mr v} >0$, \citep{1972AZh....49..334S,1984ApJ...279..882E,1987ApJ...321..721H}, which could lead to growth
of Langmuir waves.
The beam-plasma instability acts to decrease the gradient of the electron distribution in velocity space,
while Coulomb collisions tend to restore it.  The net result is a plateau-like
distribution,  with the plateau height slowly decreasing as energy is lost to the background
plasma via collisions. It has been shown \citep[e.g.][]{1987ApJ...321..721H,1987A&A...175..255M,2011A&A...529A.109H}
that even for the time independent injection of electrons into plasma, the electrons should effectively generate
Langmuir waves, however the time or space integrated electron
spectrum (to compare with the HXR observations) is only weakly affected by this process.
The overall evolution of the beam-plasma system is complicated by various non-linear processes,
which can affect the spectrum of Langmuir turbulence \citep[see][as a review]{1995lnlp.book.....T}.
Due to these processes, the energy of Langmuir waves can be transferred from one
range of phase velocities to another, hence suppressing the growth of Langmuir waves.
The removal of Langmuir waves out of resonance with the electron
beam can potentially lead to nonlinear stabilization of the beam-plasma instability
 \citep{1975PhFl...18.1769P,1979ApJ...233..717V,1985A&A...142..219R,1991PhFlB...3.1968M}.
While these earlier works were predominantly analytical estimates, the recent weakly turbulent
treatments \citep[e.g.][]{2001PhPl....8.3982Z,2002PhRvE..65f6408K,2011ApJ...727...16Z}
are self-consistent and capture a complex interplay between wave-particle
and wave-wave processes in collisionless plasma. For beam relaxation in a plasma,
these processes can lead to the appearance of electrons with velocities
above the maximum injected. At short scales,
beam-plasma interaction Vlasov or PIC simulations
\citep[e.g.][]{2009A&A...506.1437K,2010SoPh..267..393T,2011PhPl...18e2107D,2011PhPl...18e2903T}
are particularly useful to capture phase-space effects of beam-plasma interaction at short time scales.
However, the interpretation of solar flare HXR emission requires knowledge of electron evolution
at much longer time-scales, inaccessible by Vlasov or PIC approaches.

In this paper, we investigate the evolution of non-thermal electrons and Langmuir waves using weak-turbulence
theory in a collisional, inhomogeneous solar coronal plasma. We show that the evolution of the Langmuir
wave spectrum caused by plasma inhomogeneities and wave-wave processes
leads to effective acceleration of high energy electrons. The resulting HXR spectrum is found to be strongly affected
by the evolution of Langmuir turbulence, so that if the HXR spectrum is interpreted in terms of a collisional model,
the accelerated electron spectrum will be overestimated.

\section{Weak turbulent evolution of energetic electrons in a collisional plasma}

The weakly turbulent collisional relaxation of an energetic electron population is considered
in a collisional plasma typical for the solar corona. The evolution of energetic electrons is described using weak turbulence
theory including wave-particle and wave-wave interactions in non-uniform plasma. The corresponding
equations \citep{1963JETP...16..682V,1967PlPh....9..719V,1995lnlp.book.....T}
governing the electron distribution function $f({\mr v},t)$ [electrons cm$^{-3}$ (cm/s)$^{-1}$ ]
and the spectral energy densities of Langmuir $W_k$, and ion-sound waves $W_k^s$
[ergs cm$^{-2}$ ] are as in \citet{2002PhRvE..65f6408K}:
\begin{flushleft}
\begin{equation}
\frac{\partial f}{\partial t}= \frac{4\pi^2 e^2}{m^2}\frac{\partial}{\partial {\mr v}}\left(
\frac{W_k}{{\mr v}}\frac{\partial f}{\partial {\mr v}}\right) +{\rm St}_{col}(f),\;\;\; \label{eqk1}
\end{equation}
\begin{eqnarray}
\frac{\partial W_k}{\partial t}-\frac{\partial \omega_{pe}(x)}{\partial x}
\frac{\partial W_k}{\partial k} =\frac{\pi\omega_{pe}^3}{nk^2}W_k\frac{\partial f}{\partial {\mr v}}-\gamma_{col}W_k+\cr
\frac{\omega_{pe}^3 m_e}{4\pi n_e}{\mr v} \ln\left(\frac{{\mr v}}{{\mr v}_{Te}}\right) f+{\rm St}_{decay}(W_k,W^s_k).
\label{eqk2}
\end{eqnarray}
Here, ${\rm St}_{decay}(W_k,W^s_k)$ is the integral describing the parametric decay of a Langmuir wave
with an ion-sound wave represented by $l\rightarrow l^{\prime}+s$,
\begin{eqnarray}
{\rm St}_{decay}(W_k,W^s_k)=\alpha\omega_{k}\times \;\;\;\;\;\;\;\;\;\;\;\;\;\;\;\;\;
\;\;\;\;\;\;\;\;\;\;\;\;\;\;\;\;\;\;\;\;\;\;\;\;\;\;\;\;\;\;
\cr
\int\omega^s_{k^{\prime}}\left[ \left(
\frac{W_{k-k^{\prime}}}{\omega_{k-k^{\prime}}}\frac{W^s_{k'}}{\omega^s_{k'}}-
\frac{W_k}{\omega_k}\left(\frac{W_{k-k^{\prime}}}{\omega_{k-k^{\prime}}}+
\frac{W^s_{k^{\prime}}}{\omega^s_{k^{\prime}}}\right)\right)\right.\times
\cr
\delta (\omega_{k}-\omega_{k-k^{\prime}}-\omega^s_{k^{\prime}})-
\left.\left(\frac{W_{k+k^{\prime}}}{\omega_{k+k^{\prime}}}\frac{W^s_{k'}}{\omega^s_{k'}}-
\frac{W_k}{\omega_k}\left(\frac{W_{k+k^{\prime}}}{\omega_{k+k^{\prime}}}-
\frac{W^s_{k^{\prime}}}{\omega^s_{k^{\prime}}}\right)\right)\times\right.\cr
\left.\delta(\omega_{k}-\omega_{k+k^{\prime}}+\omega^s_{k^{\prime}}) \right.\biggr] \,dk^{\prime} \;
\end{eqnarray}
with
\begin{equation}
\label{alphaL}
 \alpha=\frac{\pi \omega^2_{pe}(1+3T_i/T_e)}{4n\kappa T_e}.
\end{equation}
\end{flushleft}
where the spectral energy density is normalized such, that  $\int W_k\, dk$
is the energy density of the waves [erg cm$^{-3}$].
The system of equations (\ref{eqk1}-\ref{alphaL}) describes the weakly turbulent
evolution of electrons and  Langmuir waves in the presence of density inhomogeneities
and/or ion-sound waves. The spontaneous emission of Langmuir waves has been treated
as in \citet{2009ApJ...707L..45H,2011A&A...529A..66R} and the system was numerically
integrated using a numerical scheme as in \citet{2001CoPhC.138..222K}.

As we are interested in the evolution of the system at time-scales $t\gg \tau _{col}$,
the collisional operator \citep[e.g.][]{1981phki.book.....L} is included to account
for the binary collisions
\begin{equation}\label{St_col}
{\mr St}_{col}(f)=\Gamma\frac{\partial}{\partial {\mr v}}\left(\frac{f}{{\mr v}^2}+\frac{{\mr v}_{Te}^2}{{\mr v}^3}\frac{\partial f}{\partial {\mr v}}\right),
\end{equation}
where the first RHS term describes the systematic drag on energetic particles
and the second term diffusion in velocity space due to binary collisions.
We note that the collisional drag is larger than energy losses due to spontaneous
emission of Langmuir waves [second term in the right hand side of Eq. (\ref{eqk1})]
by a non-thermal electron in a plasma. In other words, only a small part of the energy lost
by an electron with velocity ${\mr v}$ goes into spontaneously generated Langmuir waves
given by Eq.  (\ref{eqk2}). Finally, the collisional damping rate for Langmuir
waves $\gamma_{col} \simeq \Gamma/4{\mr v}_{Te}^3$ is added
to Eq. (\ref{eqk2}).

\subsection{Initial electron distribution function}
Let us consider the evolution of a non-thermal  electron population with the initial
electron distribution $g_0({\mr v})$ in a Maxwellian plasma (Fig. \ref{fig:collisions}):
\begin{equation}
f({\mr v}, t=0)= \frac{n}{\sqrt{2\pi} {\mr v}_{Te}} \exp\left(-\frac{{\mr v}^2}{2 {\mr v}_{Te}^2}\right) +g_0({\mr v}),
\label{eq:f_t0}
\end{equation}
where the non-thermal particle distribution $g_0({\mr v})$ is initially a power law $f({\mr v},t=0)\sim {\mr v}^{-2\delta}$
for ${\mr v}> {\mr v}_b=10{\mr v}_{Te}$ and flattens at low velocities ${\mr v}<10{\mr v}_{Te}$;
\begin{equation}
g_0({\mr v})=\frac{2 n_{b}}{\sqrt{\pi}\, {\mr v}_b}\frac{\Gamma(\delta)}{ \Gamma(\delta-\frac{1}{2})}
\left[1+({\mr v}/{\mr v}_{b})^2\right]^{-\delta},
\label{eq:g0}
\end{equation}
where $\delta$ is the power law index for the energetic particles in energy space,
$n_{b}$ the number density of non-thermal electrons, $n_{b}<<n$, and $\Gamma (x)$ denotes the gamma
function. The initial electron distribution is normalized to the electron number density [electrons cm$^{-3}$],
so that
\begin{equation}
 \int_{0}^\infty g_0({\mr v})\mbox{d}{\mr v}=n_{b}.
\end{equation}
The initial level of Langmuir waves can be calculated assuming an equilibrium
with Maxwellian electron distribution function, ignoring all non-linear terms
and setting $dW_k/dt=0$ in Eq. (\ref{eqk2})
\begin{eqnarray}
W(k, t=0)\simeq \frac{k_b T_e}{4 \pi^2}\frac{k^2\ln(\frac{1}{k\lambda_{de}})}
{1+\sqrt{\frac{2}{\pi}}\frac{\gamma_{col}}{\omega_{pe}}k^3\lambda_{De}^3 \exp(\frac{1}{2k^2\lambda_{De}^2})}=\cr
\frac{k_b T_e}{4 \pi^2}\frac{k^2\ln(\frac{1}{k\lambda_{de}})}
{1+\frac{\ln \Lambda}{16\pi n_e}\sqrt{\frac{2}{\pi}}k^3 \exp(\frac{1}{2k^2\lambda_{de}^2})},
\label{eq:W_t0}
\end{eqnarray}
where $k_b$ is Boltzmann constant, $T_e=m{\mr v}_{Te}^2$ the electron temperature of the plasma,
and $\lambda_{De}={\mr v}_{Te}/\omega _{pe}$ is the Debye length.
The equation (\ref{eq:W_t0}) can be reduced to $W(k, t=0)\simeq \frac{k_b T_e}{4 \pi^2}{k^2\ln(\frac{1}{k\lambda_{de}})}$
in the collisionless limit for $\gamma _{col}\rightarrow 0$,
which is the thermal level of plasma waves in a collisionless Maxwellian
plasma \citep{1973plas.book.....K,1995lnlp.book.....T}. For the problem considered,
the exact initial level of plasma waves is not important, as the governing equations quickly
establish a balanced level, which is a few orders of magnitude less
than the level driven by the instability.

\section{Evolution of the electron distribution function}

We assume a background plasma similar to the solar corona, with plasma frequency
of $f_{pe}=\omega _{pe}/2\pi =2$~GHz and an electron and ion temperature of $T_i=T_e=1$~MK.
The corresponding plasma density is then $5.0\times 10^{10} $cm$^{-3}$ and the collisional timescale
is $\tau_{col}=1/\gamma _{col} = 7\times 10^{-5}$ seconds. For the electron beam we take $n_{b}/n = 1\times 10^{-2}$,
and $\delta=4$, giving a power-law index of $8$ in velocity space above approximately ${\mr v}_{b}$.

\subsection{Collisional relaxation}

\begin{figure}
\center
\includegraphics[width=7cm]{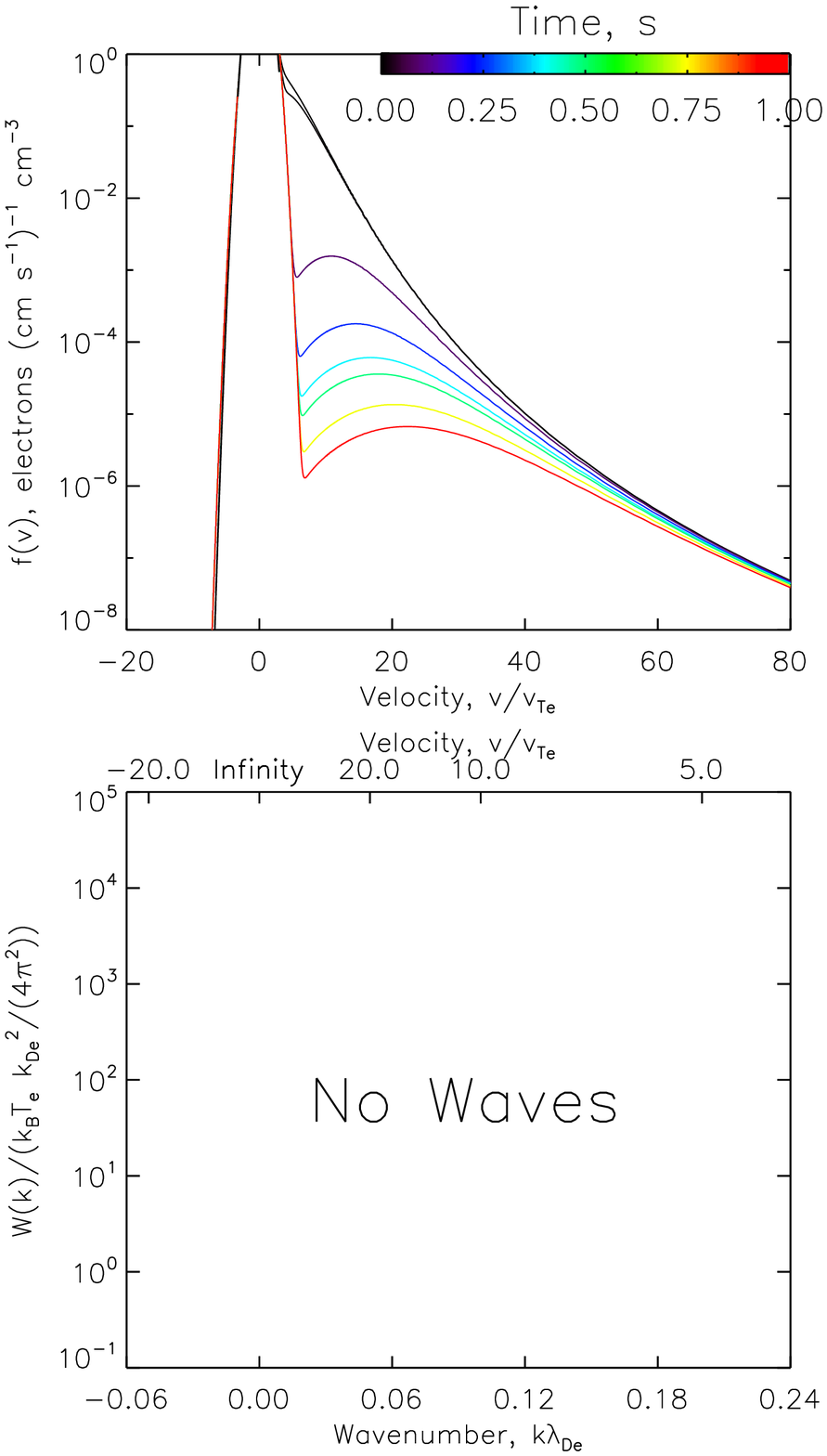}
\includegraphics[width=7cm]{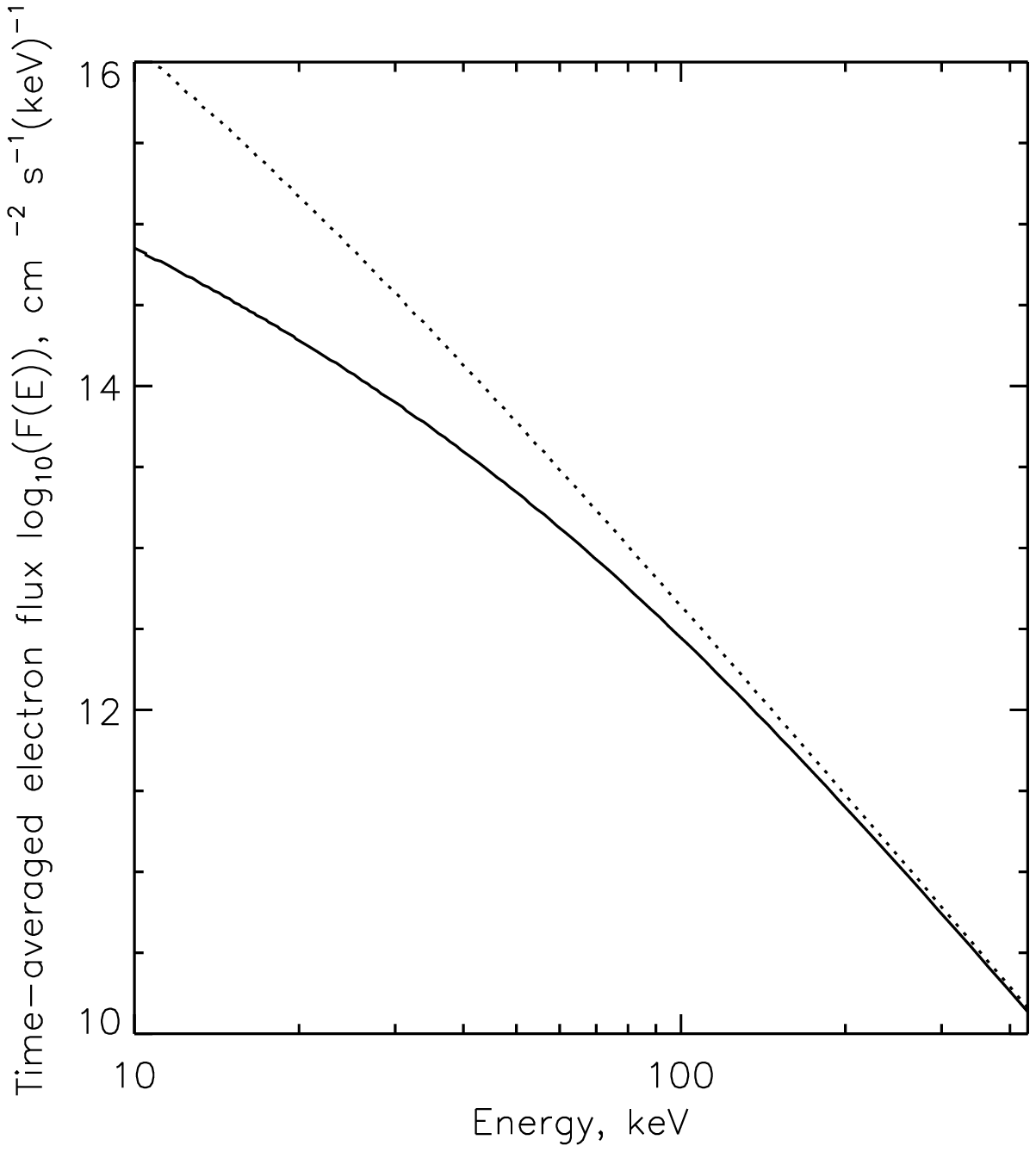}
\caption{Collisional relaxation of an electron beam in a plasma. Top and middle panels; the electron distribution function $f({\mr v})$  and the spectral energy density of Langmuir waves (here fixed at the thermal level), respectively. Each coloured line shows the distribution at a different time. Bottom panel; the time averaged electron flux spectrum [electrons keV$^{-1}$ cm$^{-2}$ s$^{-1}$] plotted against electron energy. The dashed line shows the flux if the beam remained the initial power-law.}
\label{fig:collisions}
\end{figure}

Under the influence of Coulomb collisions, the electron distribution function for ${\mr v}\gg {\mr v}_{Te}$
evolves according to
\begin{equation}\label{Eq:St_col}
\frac{\partial f}{\partial t}= -\frac{\partial}{\partial {\mr v}}\left(\frac{d{\mr v}}{dt}f\right)=\frac{\partial}{\partial {\mr v}}\left(\Gamma\frac{f}{{\mr v}^2}\right),
\end{equation}
which can be straightforwardly solved for arbitrary initial distribution function, $g_0({\mr v})$
so that $f({\mr v},t)=[{\mr v}/u({\mr v},t)]^2g_0(u({\mr v},t))$, where $u({\mr v}, t)=({\mr v}^3+3\Gamma t)^{1/3}$.
The time integrated, or mean electron distribution, will be the so-called
thick-target electron spectrum. This expression is, as expected, similar to the space integrated
collisional thick-target model \citep[e.g.][]{1963SPhD....8..543D,1971SoPh...18..489B,2003ApJ...595L.115B}
and directly corresponds to the electron spectrum responsible for X-ray flux.
Figure (\ref{fig:collisions}) shows the time evolution of the electron
distribution and the time integrated electron flux spectrum.
The time-dependent positive slope $\partial f/\partial {\mr v}>0$ is formed
for velocities ${\mr v}<(3\Gamma t)^{1/3}$
while the tail at ${\mr v}>(3\Gamma t)^{1/3}$ is weakly affected by collisions.
As time progresses, a larger range of electron velocities shows a distribution function which grows
with velocity.

\subsection{Langmuir wave generation in uniform plasma}

\begin{figure}
\center
\includegraphics[width=7cm]{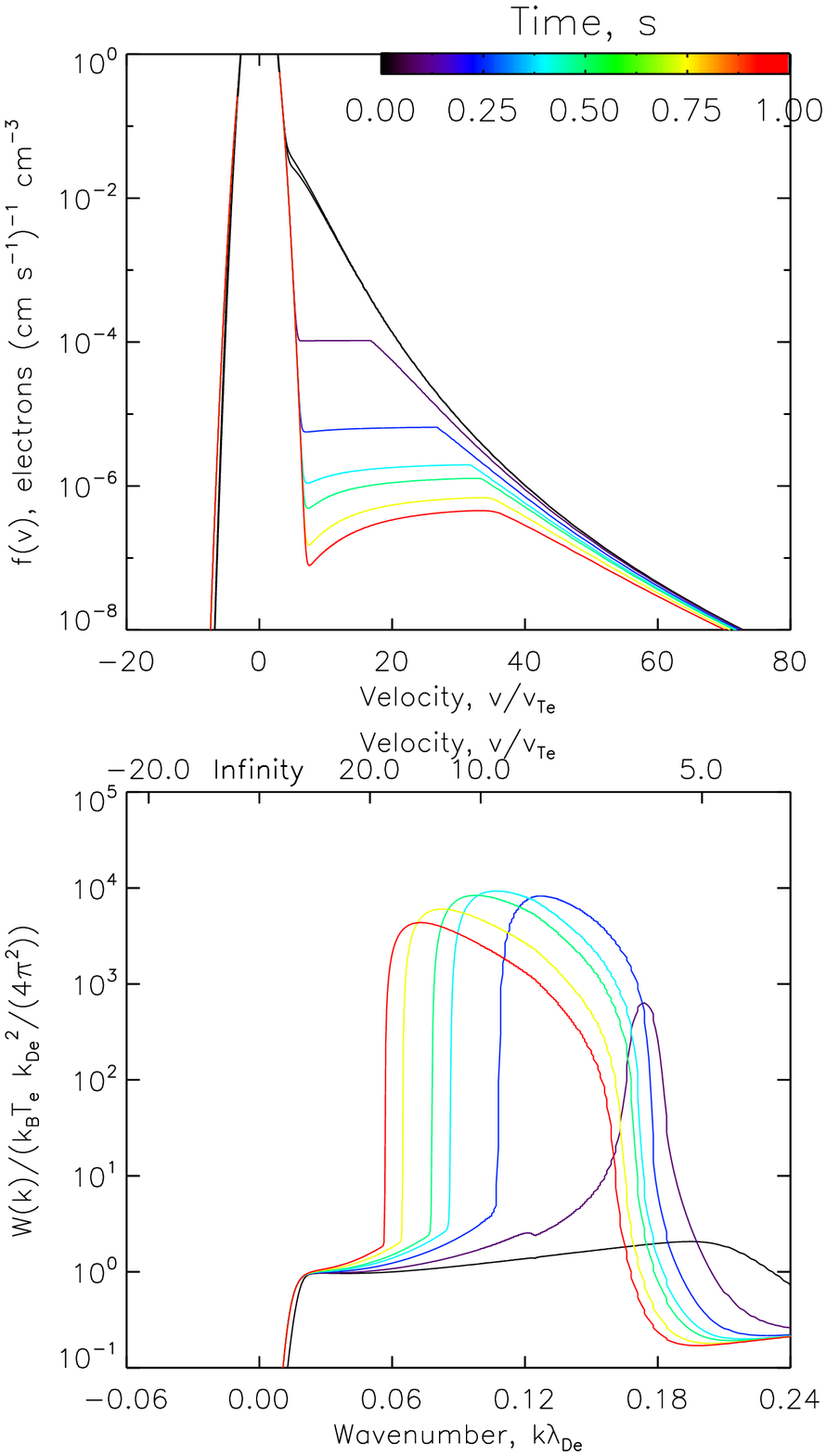}
\includegraphics[width=7cm]{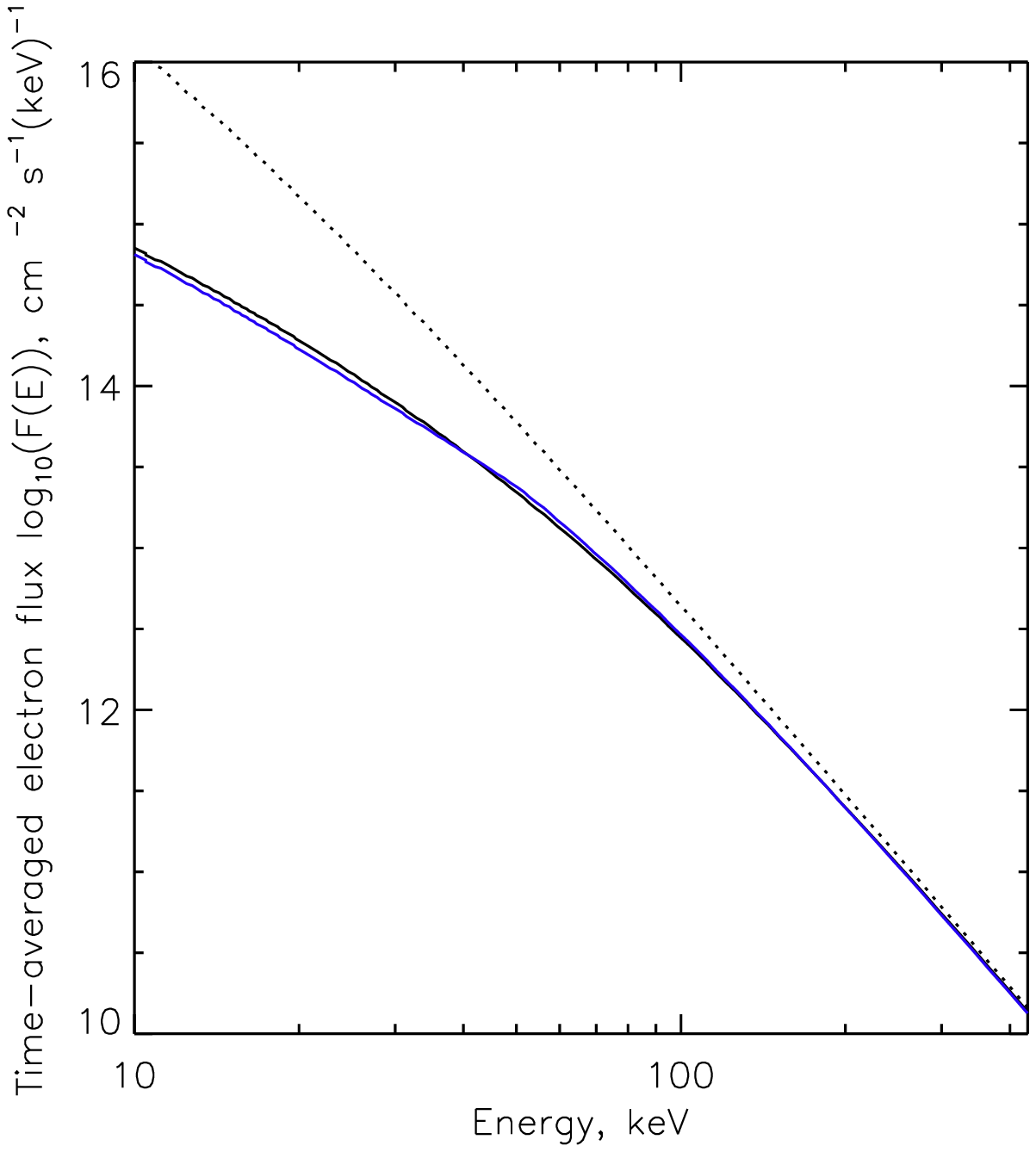}
\caption{The same as in Figure \ref{fig:collisions} but for collisional relaxation including
Langmuir wave generation in homogeneous plasma. In the bottom panel, the black line shows the collisions only case,
the blue included wave generation.}
\label{fig:langmuir}
\end{figure}

In the case of a uniform plasma ($\partial {\omega _{pe}}/{\partial x}=0$)
and for simplicity ignoring all wave-wave terms, the collisionally formed unstable distribution
will lead to instability and effective generation of plasma waves (Fig. \ref{fig:langmuir}).
The generation of Langmuir waves quickly flattens the distribution function
and forms a plateau between $\sim 3{\mr v}_{Te}<{\mr v}<(3\Gamma t)^{1/3}$ (Fig. \ref{fig:langmuir}).
Over time, more electrons are lost due to collisions, and the plateau in velocity space becomes
wider. As soon as the beam density has dropped, Langmuir waves are no longer generated appreciably,
and the evolution of the electron beam becomes again purely collisional. The important aspect
is that while the instantaneous distribution of electrons in the cases of quasi-linear and collisions-only
relaxation are very different, the time-integrated spectrum of energetic electrons is almost identical.
This confirms the previously published results
\citep{1987ApJ...321..721H,1987A&A...175..255M,2009ApJ...707L..45H,2011A&A...529A.109H} that the quasi-linear interaction
weakly affects the time-integrated spectrum. The energy is quickly transferred from electrons
to the plasma waves and back, the time integrated amount of energy for a given resonant
velocity is almost constant \citep[e.g.][]{1963JETP...16..682V,1999SoPh..184..353M}.
The collisional energy loss of Langmuir waves does not facilitate the transfer of energy
between different velocities, but simply reduces the energy.

\subsection{Spectral evolution of Langmuir turbulence}

The Langmuir wave  spectrum can quickly evolve in $k$ space,
changing the wave energy available at different resonant
velocities, ${\mr v}=\omega _{pe}/k$. Firstly, the nonlinear wave-wave interactions
change the distribution of Langmuir wave energy \citep{1967PlPh....9..719V,1974ApJ...190..175P,1982PhFl...25.1062G,2000A&A...353..757B,2005PhRvL..95u5003Y,2008PhRvL.100e5006F,2009PhPl...16i2304P}
and secondly, the waves can be effectively scattered
or refracted by inhomogeneous plasma
\citep{1969JETP...30..131R,1969JETP...30..759B,1978SvJPP...4R1267K,1975PhFl...18..679C,1976JPSJ...41.1757N,1979ApJ...233..998S,1980SvJPP...6Q.137A,1989SoPh..123..343M,
2001SoPh..202..131K,2001A&A...375..629K,2006PhPl...13i2902L,2011ApJ...727...16Z}.

\subsubsection{Inhomogeneous plasma - constant density gradient}

To illustrate the role of the spectral evolution of Langmuir waves on the electron distribution function, we consider the relaxation
of energetic electrons in a plasma with a constant density gradient, so that the term describing refraction
of Langmuir waves is
\begin{equation}\label{eq:omega_L}
\frac{\partial \omega_{pe}}{\partial x}=\frac{\omega_{pe}}{L}
\end{equation}
where \begin{equation}\label{L}
L\equiv \omega_{pe}(x)\left(\frac{\partial \omega _{pe}(x)}{\partial x} \right)^{-1}
=\frac{n_{pe}(x)}{2}\left(\frac{\partial n_{pe}(x)}{\partial x} \right)^{-1}
\end{equation}
is the characteristic scale of density inhomogeneity. Langmuir waves
propagating into the region of higher/lower density experience change in wavenumber
to  $k+\Delta k$, with $\Delta k \simeq \pm \omega_{pe}\Delta t/|L|$,
depending on the sign of the density gradient. This in turn affects the
resonant velocity ${\mr v}=\omega_{pe}/k$, so electrons of smaller/larger velocities
can now interact with this wave.
\begin{figure}
\centering
\includegraphics[width=7cm]{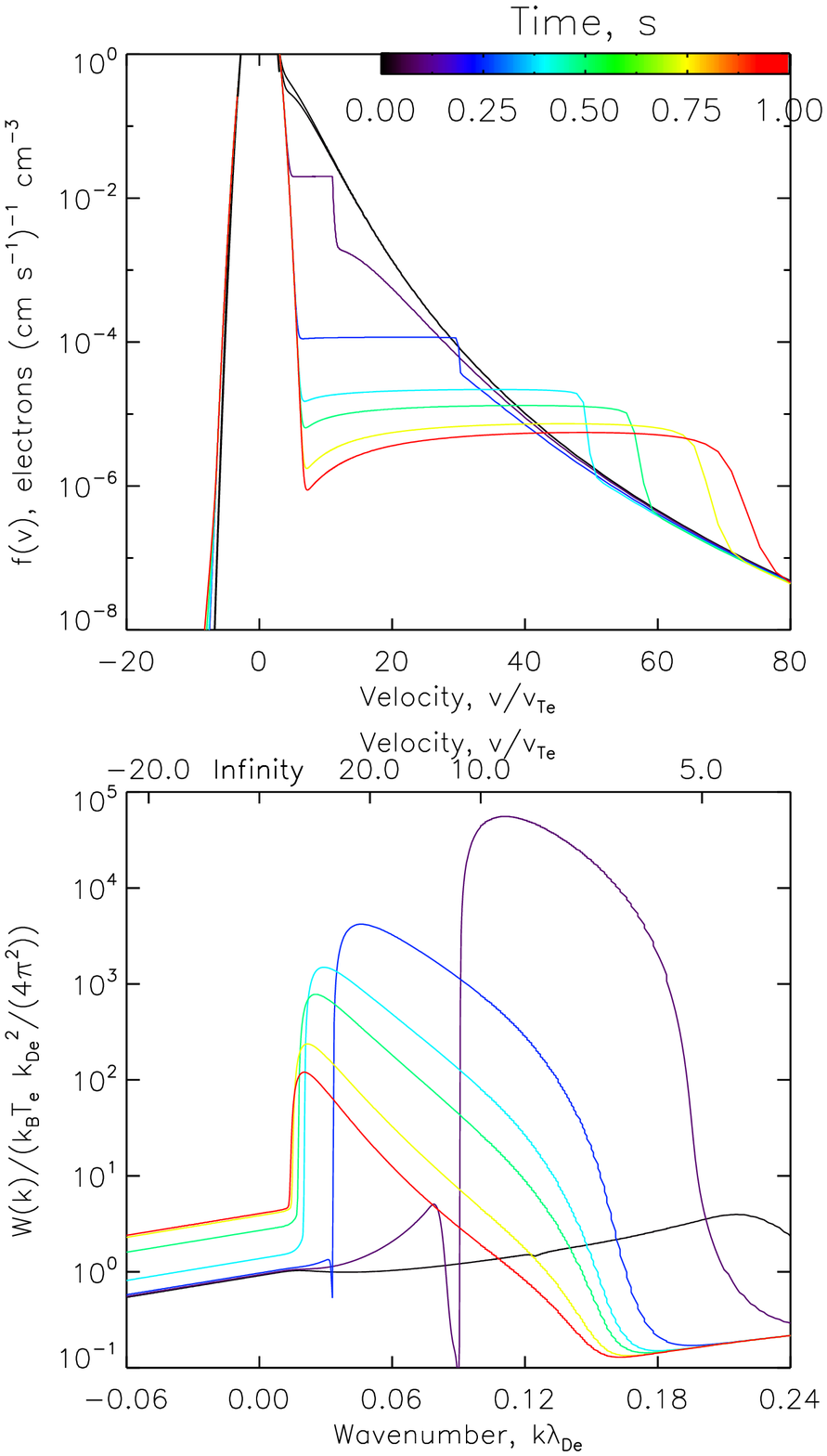}\\
\includegraphics[width=7cm]{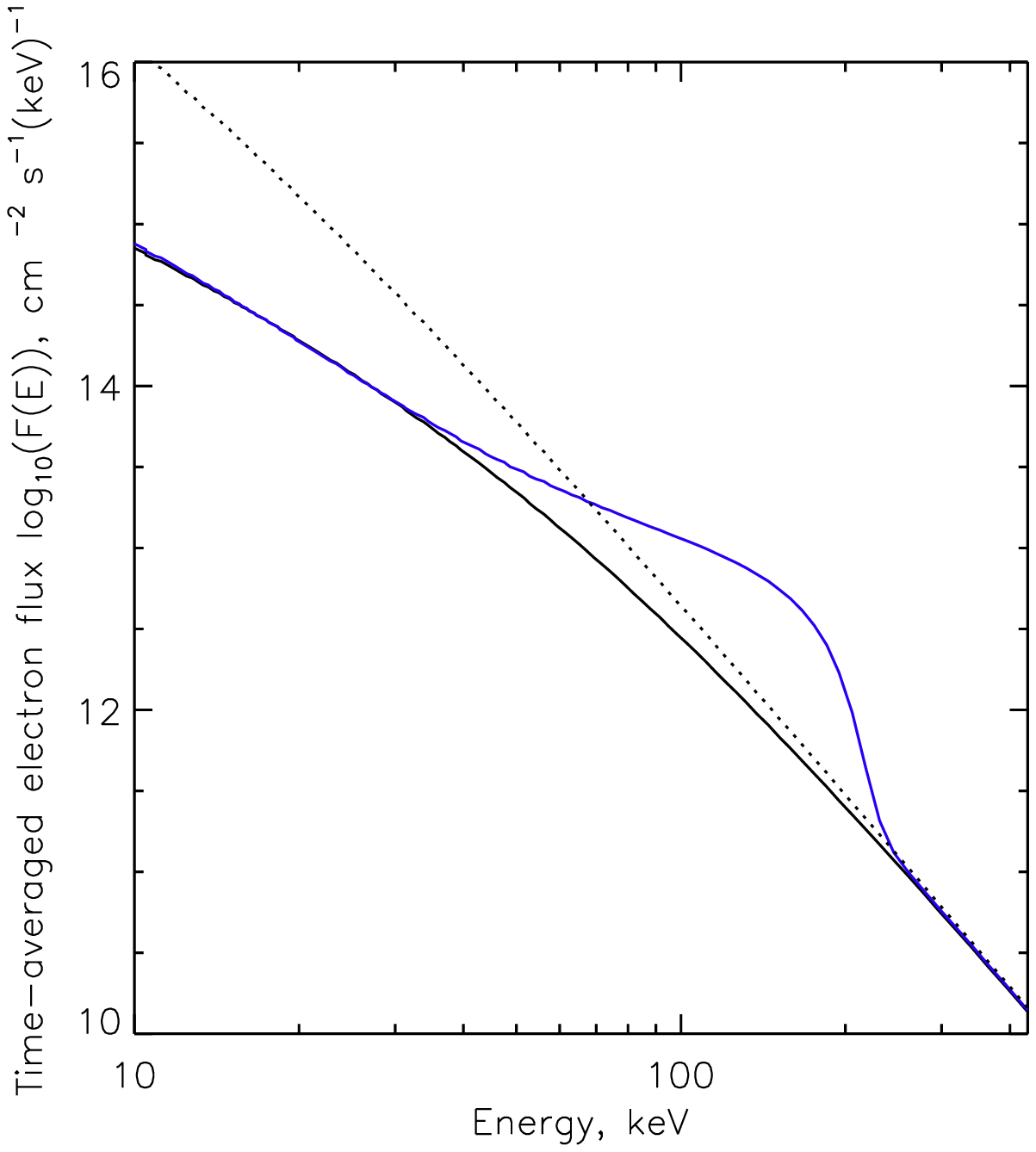}
\caption{The same as in Figure \ref{fig:langmuir}, but for a plasma with constant density
gradient as in Equation \ref{eq:omega_L}, with $L=10^6$~cm. } \label{fig:Grad}
\end{figure}
For electrons propagating into denser regions, the plasma density grows, causing
generated Langmuir waves to have a negative shift in $k$, $\Delta k \simeq- \omega_{pe} \Delta t/|L|$.
Therefore, Langmuir waves generated at larger $k$ are shifted towards
smaller $k$ (higher phase velocities),  and hence, can be effectively absorbed by faster electrons
with ${\partial f}/{\partial {\mr v}}<0$, which are consequently accelerated.
For a density gradient of $L\gtrsim 10^6$~cm, the waves are shifted
faster than the characteristic time of collisional relaxation, so for $L\sim {\mr v}_{Te}\gamma _{col}^{-1} \sim 10^6$~cm,
substantial energy can be transferred to higher electron velocities. Figure \ref{fig:Grad} shows
clear acceleration of energetic electrons above $\sim 30 {\mr v}_{Te}$ as the Langmuir waves shift
to higher phase velocities. Importantly, these accelerated electrons
are evident in the time averaged spectrum, hence the resulting Hard X-ray spectrum will be affected
and the role of Langmuir waves is detectable.

\subsubsection{Density fluctuations}

\begin{figure}
\centering
\includegraphics[width=7cm]{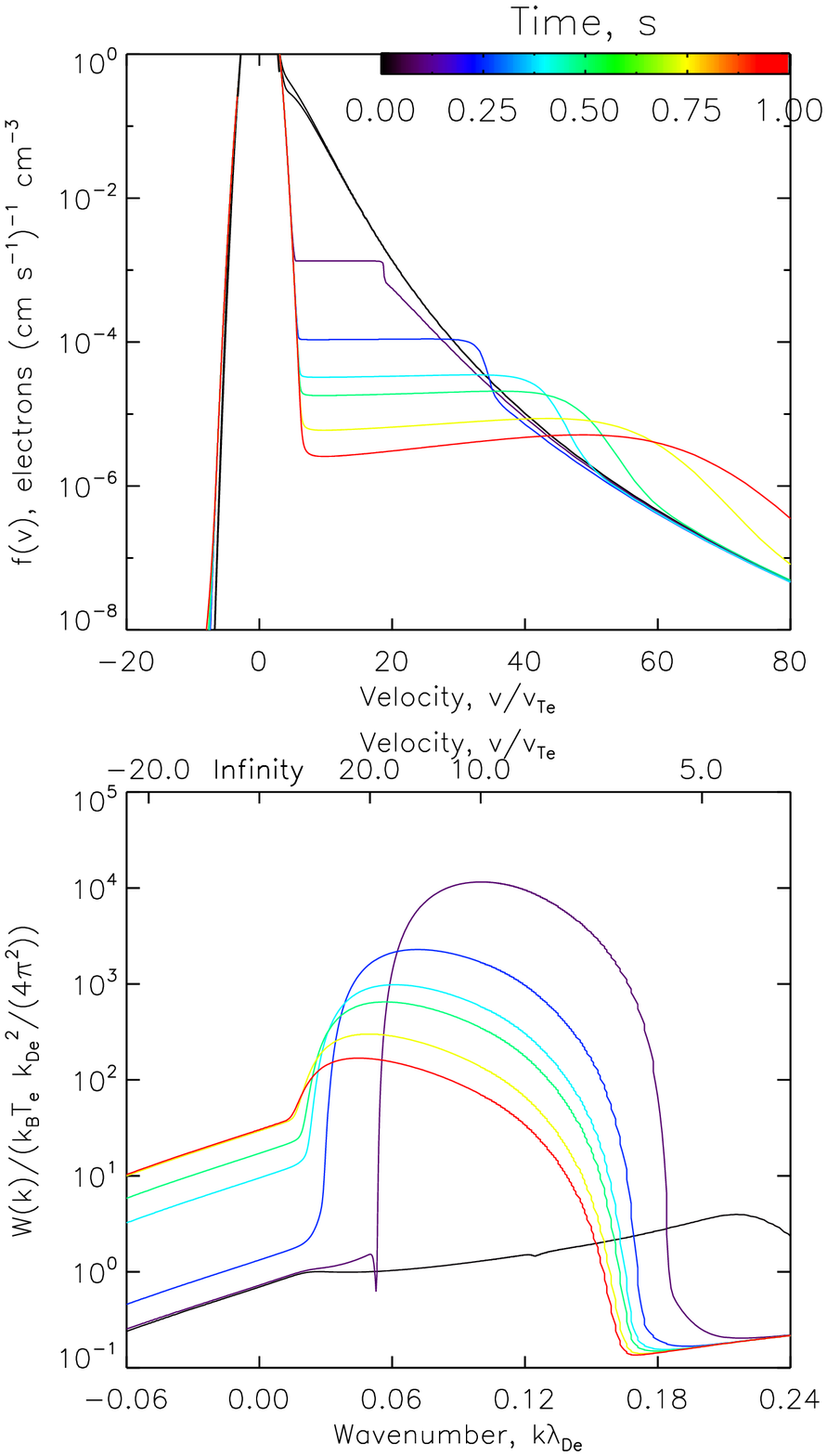}\\
\includegraphics[width=7cm]{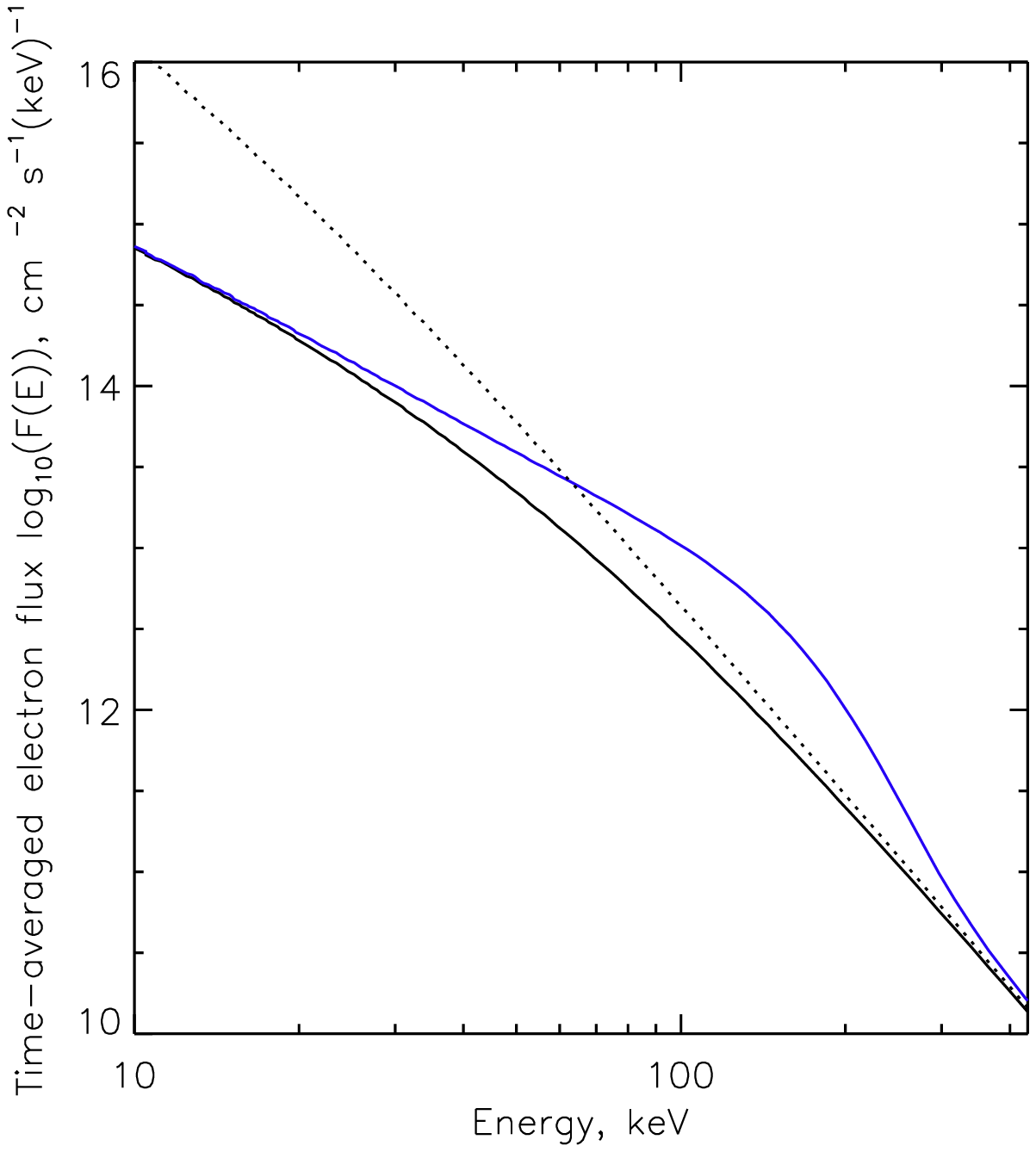}
\caption{The same as in Figure \ref{fig:langmuir}, but for the case with random density
fluctuations, given by Equation (\ref{eq:Diff_k}).}
\label{fig:diffusion}
\end{figure}

However, even in the absence of a constant density gradient, the Langmuir wave spectrum
is also expected to evolve. For instance, the solar coronal plasma, as any real plasma, has a fluctuating
part, $n+\delta n(x,t)$. For large scale density perturbations $1/L<<\omega _{pe}/{\mr v}_b$, the resonant conditions for Langmuir
wave decay or scattering cannot be satisfied and the Langmuir waves will be scattered with a small
change in $k$ \citep{1976JPSJ...41.1757N,1982PhFl...25.1062G}. Let us consider
density fluctuations with $\langle(\delta n/n)^2\rangle<<1$, zero mean $\langle\delta n\rangle=0$
and a Gaussian spectrum $|\delta n^2|_{\Omega, q}\sim \exp(-q^2/q_0^2)\exp(-\Omega^2/\Omega_0^2)$,
where $q_0$ and $\Omega_0$ are their characteristic wavenumber and frequency.
Under the influence of these turbulent density fluctuations,
the spectral evolution of Langmuir waves in $k$ becomes diffusive
\begin{equation}
-\langle\frac{\partial \omega_{pe}}{\partial x}\frac{\partial W_k}{\partial k}\rangle\rightarrow \frac{\partial}{\partial k} \left(D_k \frac{\partial W_k}{\partial k}\right),
\end{equation}
where $\langle\rangle$ denotes ensemble average over density fluctuations. The diffusion coefficient describing this
random refraction of waves is
\begin{equation}
D_{k}=\frac{\sqrt{\pi}}{4} \omega_{pe}^2\langle({\delta n}/n)^2\rangle \frac{q_0}{u_0} \left(1+\frac{k^2}{k_\star^2}\right)^{-3/2},
\label{eq:Diff_k}
\end{equation}
where $k_\star={2u_0\omega_{pe}}/{3 {\mr v}_{Te}^2}$ and $u_0=\Omega_0/q_0$.
Assuming a modest level of turbulence ${\delta n}/n= 10^{-3}$,
with a characteristic speed close to the speed of sound in the corona $u_0=\Omega_0/q_0\sim 10^7$~cm~s$^{-1}$
and long wavelength $1/q_0 \sim 10^{4}$~cm, this spetral diffusion $D_{k}$
substantially changes the spectrum of Langmuir waves in a few collisional times.
Figure \ref{fig:diffusion} shows the evolution of the electron distribution function and
the spectral energy density of Langmuir waves. Langmuir waves diffuse in $k$-space,
from an initial value of $k\sim\omega_{pe}/{\mr v}_b$,
so in time the spectrum of Langmuir waves becomes broader as time increases.
The absorption of Langmuir waves by electrons above $20 {\mr v}_{Te}$
leads to electron acceleration. As in the constant gradient case, the integrated electron
flux for the fluctuating density case changes considerably at energies above $\sim 20$~keV.

\subsubsection{Three-wave interaction of Langmuir waves}
For density fluctuations with wavenumber $q \sim \omega_{pe}/{\mr v}_b$ the diffusive treatment is no longer valid,
and we must individually consider the interactions between Langmuir waves and other modes in the plasma.
In particular, we are interested in those modes which may be excited by the high level of Langmuir waves present
due to the beam-plasma interaction.
Following the treatments in uniform \citep{2001PhPl....8.3982Z}
and inhomogeneous \citep{2002PhRvE..65f6408K,2011ApJ...727...16Z} plasmas,
we now consider the three wave process, $ l^{\prime}$; $l\rightarrow l^{\prime}+s$,
where $l$ and $l'$ are Langmuir waves and $s$ is a sound wave.
This process efficiently scatters electron beam generated waves with wavenumber ${\bf k}$,
into secondary Langmuir waves with wavenumber ${\bf k}^{\prime}\approx {\bf -k}$.
The decay of a Langmuir wave
generates a ion-sound wave with ${\bf q}\approx 2{\bf k}$. Every scattering decreases
the absolute value of the Langmuir wave number
by $k^{*}_d=2\sqrt{m_e/m_i}\sqrt{1+3T_i/T_e}/(3\lambda_{De})$.
Hence, repeated wave-wave processes cause Langmuir waves to appear
at lower $k$ (larger phase velocities). As in the previous cases,
Langmuir waves can be absorbed by the tail of the distribution,
leading to acceleration of non-thermal electrons.

The simultaneous evolution of both ion-sound waves and Langmuir waves
has been already simulated \citep[e.g.]{2001PhPl....8.3982Z,2002PhRvE..65f6408K,2011ApJ...727...16Z}
and does produce the effect outlined in the previous paragraph. Here we limit ourself
to a more simple but quite illuminating situation with a fixed level of density fluctuations
produced by ion-sound waves. We assume the level of ion-sound waves to be thermal,
\[W_k^s= k_B T_e k_{De}^2 \frac{k_{De}^2}{k_{De}^2+k^2} \] \begin{figure}.
\centering
\includegraphics[width=7cm]{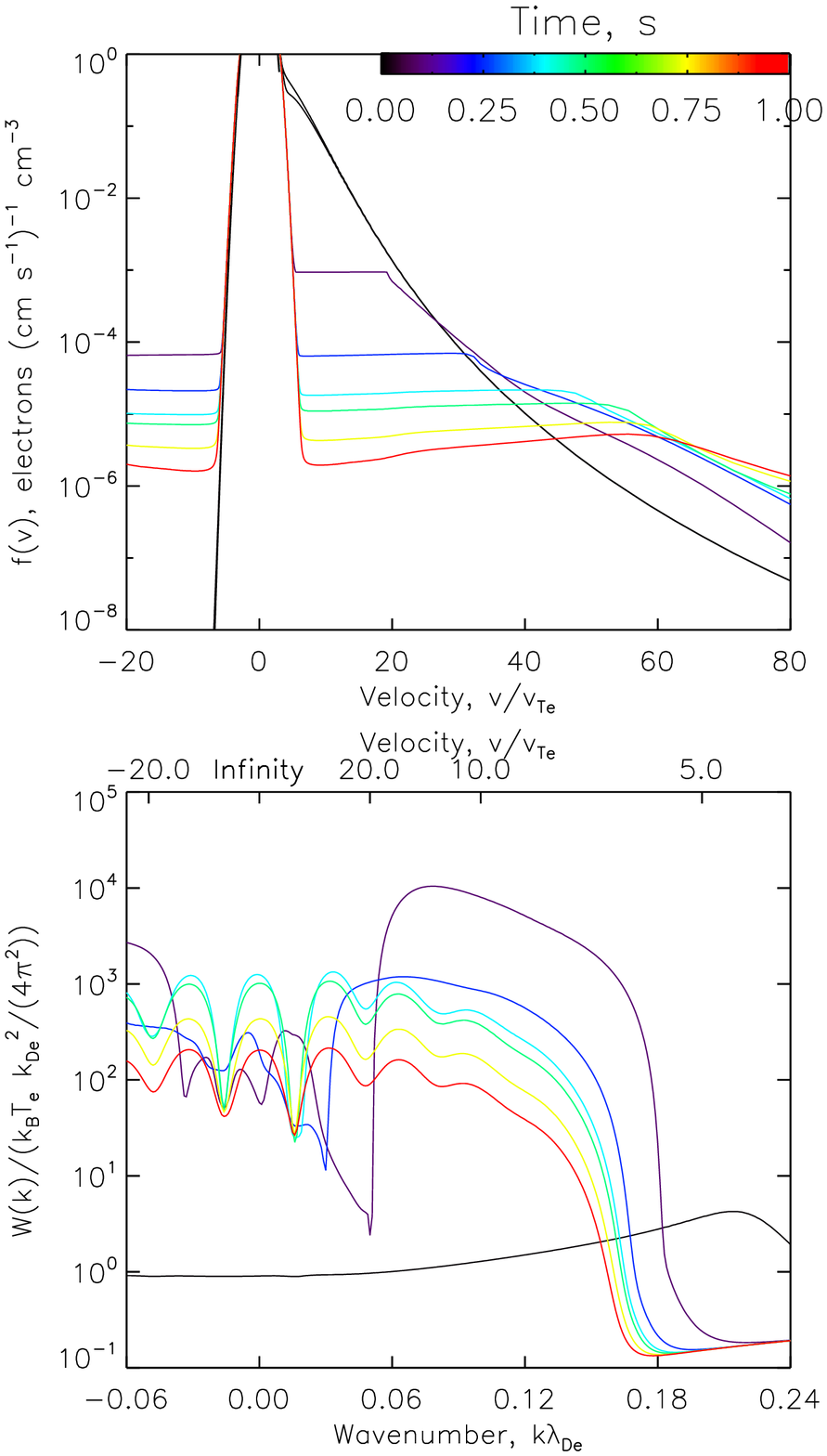}\\
\includegraphics[width=7cm]{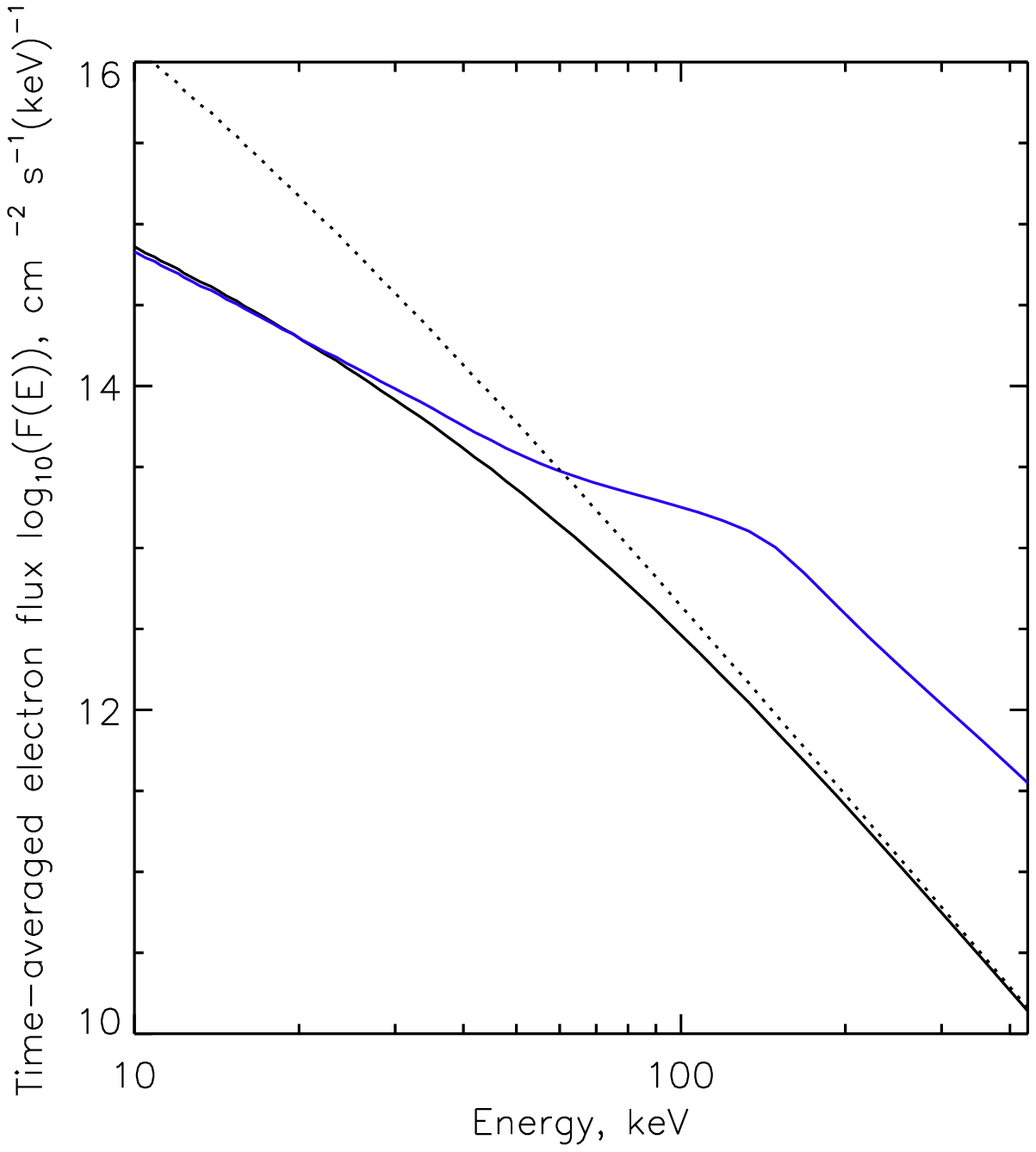}
\caption{The same as Figure \ref{fig:langmuir} but with ion-sound wave interactions as well as density fluctuations.}
\label{fig:random_nonlin}
\end{figure}
The results are presented in Fig. \ref{fig:random_nonlin}.  As in the cases of constant
density gradient and random density fluctuations, we see a time evolving plateau distribution
up to $\sim 60{\mr v}_{Te}$ at $t=1$s. The nonlinear decay term has a strong impact
on the distribution above $50{\mr v}_{Te}$ due to the high level
of waves for $k<k_{De}/60$. As the wave-wave terms are particularly effective
in shifting the Langmuir wave energy to low $k$, making Langmuir waves more isotropic,
the number of accelerated electrons at energies above $100$ keV also increases.
Below $100$ keV, the effect in the time integrated spectrum is similar
to the previous cases.

\section{Discussion and conclusions}

The above results show the significant effect that Langmuir waves have on the corresponding electron
distribution function. A constant density gradient, or random long-wavelength density fluctuations,
produce enhancement of the electron distribution in the range 30-60 ${\mr v}_{Te}$. On the other
hand, the wave-wave interactions with short wavelength fluctuations $q \sim 2k$
show a more pronounced effect at velocities ${\mr v}\simeq 60{\mr v}_{Te}$.
Unlike the case of a uniform plasma, the time averaged electron spectrum is noticeably
different from collisional relaxation. As the time averaged spectrum
can be directly compared to Hard X-ray observations, the role of Langmuir waves could be
observable.
The resulting HXR spectrum above $\sim 20$ keV is higher
than expected from collisional relaxation, and if interpreted as thick-target relaxation, the number
of energetic electrons injected is overestimated.

 \begin{figure}
\centering
\includegraphics[width=7cm]{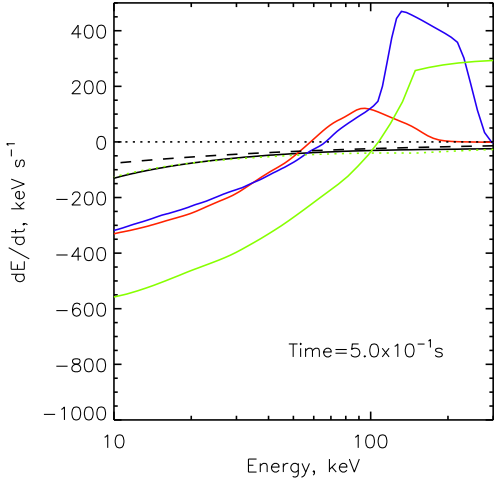}
\caption{The effective energy loss rate of an electron $\langle{\partial E}/{\partial t}\rangle_{eff}$
using Equation (\ref{eq:dE_dt}), at 0.5s. Collisional losses  $\langle{\partial E}/{\partial t}\rangle_{eff} =-\Gamma\sqrt{m^3/2}E^{-1/2}$
 from Equation \ref{dedx_TT} (black dashed line), collisional losses as in Figure \ref{fig:collisions} (black line);
Constant density gradient as in Figure \ref{fig:Grad} (blue line); Random density
fluctuations as in Figure \ref{fig:diffusion} (red line); Non-linear interactions
in inhomogeneous plasma as in Figure \ref{fig:random_nonlin} (green line).}
\label{fig:dE_dt}
\end{figure}
Given the evolution of the electron distribution, $f({\mr v},t)$, we can calculate the effective energy change rate.
Evidently the actual energy loss/gain rate is given by the kinetic equations (\ref{eqk1}-\ref{alphaL}),
but {\it assuming} a single particle evolution often used in the literature \citep[e.g.][]{2009A&A...508..993B}
the effective energy loss rate $\langle dE/dt\rangle_{eff}$ can be defined as the one satisfying a continuity
equation:
\begin{equation}\label{eq:continuity}
    \frac{\partial f(E,t)}{\partial t}+\frac{\partial }{\partial E}\left(\langle\frac{d E}{d t}\rangle_{eff}f(E,t)\right)=0
\end{equation}
where $f(E,t)=m{\mr v}(E)f({\mr v}(E),t)$ is the energy distribution function.
The effective energy loss/gain rate satisfying this equation is
\begin{equation}\label{eq:dE_dt}
    \langle\frac{d E}{d t}\rangle_{eff}=\frac{1}{f(E,t)}\int_{E}^{\infty}\frac{\partial f(E',t)}{\partial t}dE'.
\end{equation}
If the evolution of particles is governed by collisional only, the effective loss rate is simply
$\langle{\partial E}/{\partial t}\rangle_{eff} =-\Gamma\sqrt{m^3/(2E)}$ (Fig. \ref{fig:dE_dt}).
The evolving Langmuir wave turbulence notably changes  $\langle{\partial E}/{\partial t}\rangle_{eff}$,
decreasing the effective energy loss at energies of a few keV, increasing the loss rate in the intermediate
range near 10 keV, and leading to acceleration and/or reduced energy loss rate at tens of keV and above.
This effective energy change rate increases the flux of electrons above 20 keV by a factor
of a few to a few tens (shown in Figures \ref{fig:Grad}-\ref{fig:random_nonlin}).

We emphasize that the acceleration of energetic electrons is due to spectral
evolution of Langmuir waves. The Langmuir wave generation in a uniform plasma only weakly
changes the time integrated HXR spectrum and, hence, is not likely to be detected in
current HXR measurements. The important exception is the generation of waves
by spatially localized electron beams \citep[e.g.][]{2011A&A...529A.109H}, when electrons
are accelerated in bunches rather as a continuous stream as expected in some
acceleration models. The acceleration of energetic electrons
in non-uniform plasma is due to Langmuir waves generated by the non-thermal electrons
themselves and does not increase the total energy of the beam-plasma system.
On the contrary, the energetic electrons lose their energy not only via collisions but also via generation
of Langmuir waves and their subsequent absorbtion via collisions. Therefore, although the low energy electrons lose their energy faster,
the higher energy electrons are gaining energy and hence can generate HXR emission for a longer time.
This ``self-acceleration'' of some electron population reduces the number of accelerated electrons
necessary to explain the observed HXR fluxes. This process also produces a similar effect to a re-acceleration
from externally applied electric fields as discussed by \citet{2009A&A...508..993B},
but evidently does not require additional energy input.

A few important comments are due on the application of these results
to the interpretation of solar flare X-ray spectrum. The overall effect of electron re-acceleration
due to evolution of Langmuir turbulence is beam-density dependent. The higher the beam density,
the stronger the generation of Langmuir waves, and therefore the effects
considered should be more evident in HXR data. Therefore, the stronger the flare, the larger the
overestimation of non-thermal electrons, when interpreted by a collisional thick-target model.
Further, these simulations need to be extended to account for the spatial transport of energetic
electrons in solar flare conditions, a relativistic treatment of the electrons included,
and the role of whistlers \citep[e.g.][]{2002SoPh..211..135S} and kinetic Alfven
waves \citep[e.g.][]{2010A&A...519A.114B} considered.
The relaxation produces an enhanced level of Langmuir waves, which could lead to plasma microwave
emission \citep{1984ApJ...279..882E,1987ApJ...321..721H} and needs to be investigated further.
However, as the spectrum of Langmuir turbulence evolves with time, the wave-wave
and density gradient effects need to be included.

In summary, the evolving Langmuir turbulence spectrum has a noticeable
effect on both the instantaneous and time-averaged distribution functions leading
to the overestimation of electron flux (and hence energy)
in non-thermal electrons $>20$~keV,  if interpreted in terms of a standard collisional thick-target model.
Therefore, if Langmuir waves are effectively generated in flares and their spectrum
evolves, the number of accelerated electrons could be substantially
smaller than inferred from HXR thick-target interpretation.

\begin{acknowledgements}

This work is supported by a STFC rolling grant (EPK).
Financial support by the European Commission
through the HESPE Network is gratefully acknowledged.

\end{acknowledgements}

\bibliographystyle{aa}
\bibliography{refs}

\begin{thebibliography}{62}
\expandafter\ifx\csname natexlab\endcsname\relax\def\natexlab#1{#1}\fi

\bibitem[{{Asadullin} {et~al.}(1980){Asadullin}, {Batanov}, {Veriaev},
  {Sapozhnikov}, {Sarksian}, \& {Shelobkov}}]{1980SvJPP...6Q.137A}
{Asadullin}, F.~F., {Batanov}, G.~M., {Veriaev}, A.~A., {et~al.} 1980, Soviet
  Journal of Plasma Physics, 6, 137

\bibitem[{{Aschwanden}(2002)}]{2002SSRv..101....1A}
{Aschwanden}, M.~J. 2002, \ssr, 101, 1

\bibitem[{{B{\'a}rta} \& {Karlick{\'y}}(2000)}]{2000A&A...353..757B}
{B{\'a}rta}, M. \& {Karlick{\'y}}, M. 2000, \aap, 353, 757

\bibitem[{{Battaglia} \& {Kontar}(2011)}]{2011A&A...533L...2B}
{Battaglia}, M. \& {Kontar}, E.~P. 2011, \aap, 533, L2

\bibitem[{{Battaglia} {et~al.}(2011){Battaglia}, {Kontar}, \&
  {Hannah}}]{2011A&A...526A...3B}
{Battaglia}, M., {Kontar}, E.~P., \& {Hannah}, I.~G. 2011, \aap, 526, A3+

\bibitem[{{Bian} {et~al.}(2010){Bian}, {Kontar}, \&
  {Brown}}]{2010A&A...519A.114B}
{Bian}, N.~H., {Kontar}, E.~P., \& {Brown}, J.~C. 2010, \aap, 519, A114+

\bibitem[{{Bre{\v i}zman} \& {Ruytov}(1969)}]{1969JETP...30..759B}
{Bre{\v i}zman}, B.~N. \& {Ruytov}, D.~D. 1969, Soviet Journal of Experimental
  and Theoretical Physics, 30, 759

\bibitem[{{Brown}(1971)}]{1971SoPh...18..489B}
{Brown}, J.~C. 1971, \solphys, 18, 489

\bibitem[{{Brown} {et~al.}(2003){Brown}, {Emslie}, \&
  {Kontar}}]{2003ApJ...595L.115B}
{Brown}, J.~C., {Emslie}, A.~G., \& {Kontar}, E.~P. 2003, \apjl, 595, L115

\bibitem[{{Brown} {et~al.}(1990){Brown}, {Karlicky}, {MacKinnon}, \& {van den
  Oord}}]{1990ApJS...73..343B}
{Brown}, J.~C., {Karlicky}, M., {MacKinnon}, A.~L., \& {van den Oord}, G.~H.~J.
  1990, \apjs, 73, 343

\bibitem[{{Brown} {et~al.}(2009){Brown}, {Turkmani}, {Kontar}, {MacKinnon}, \&
  {Vlahos}}]{2009A&A...508..993B}
{Brown}, J.~C., {Turkmani}, R., {Kontar}, E.~P., {MacKinnon}, A.~L., \&
  {Vlahos}, L. 2009, A\&A, 508, 993

\bibitem[{{Christe} {et~al.}(2011){Christe}, {Krucker}, \&
  {Saint-Hilaire}}]{2011SoPh..270..493C}
{Christe}, S., {Krucker}, S., \& {Saint-Hilaire}, P. 2011, \solphys, 270, 493

\bibitem[{{Coste} {et~al.}(1975){Coste}, {Reinisch}, {Montes}, \&
  {Silevitch}}]{1975PhFl...18..679C}
{Coste}, J., {Reinisch}, G., {Montes}, C., \& {Silevitch}, M.~B. 1975, Physics
  of Fluids, 18, 679

\bibitem[{{Daldorff} {et~al.}(2011){Daldorff}, {P{\'e}cseli}, {Trulsen},
  {Ulriksen}, {Eliasson}, \& {Stenflo}}]{2011PhPl...18e2107D}
{Daldorff}, L.~K.~S., {P{\'e}cseli}, H.~L., {Trulsen}, J.~K., {et~al.} 2011,
  Physics of Plasmas, 18, 052107

\bibitem[{{Dubov}(1963)}]{1963SPhD....8..543D}
{Dubov}, {\'E}.~E. 1963, Soviet Physics Doklady, 8, 543

\bibitem[{{Emslie} \& {Smith}(1984)}]{1984ApJ...279..882E}
{Emslie}, A.~G. \& {Smith}, D.~F. 1984, \apj, 279, 882

\bibitem[{{Fleishman} {et~al.}(2011){Fleishman}, {Kontar}, {Nita}, \&
  {Gary}}]{2011ApJ...731L..19F}
{Fleishman}, G.~D., {Kontar}, E.~P., {Nita}, G.~M., \& {Gary}, D.~E. 2011,
  \apjl, 731, L19+

\bibitem[{{Fouquet} \& {Pesme}(2008)}]{2008PhRvL.100e5006F}
{Fouquet}, T. \& {Pesme}, D. 2008, Physical Review Letters, 100, 055006

\bibitem[{{Goldman} \& {Dubois}(1982)}]{1982PhFl...25.1062G}
{Goldman}, M.~V. \& {Dubois}, D.~F. 1982, Physics of Fluids, 25, 1062

\bibitem[{{Hamilton} \& {Petrosian}(1987)}]{1987ApJ...321..721H}
{Hamilton}, R.~J. \& {Petrosian}, V. 1987, \apj, 321, 721

\bibitem[{{Hannah} \& {Kontar}(2011)}]{2011A&A...529A.109H}
{Hannah}, I.~G. \& {Kontar}, E.~P. 2011, \aap, 529, A109+

\bibitem[{{Hannah} {et~al.}(2009){Hannah}, {Kontar}, \&
  {Sirenko}}]{2009ApJ...707L..45H}
{Hannah}, I.~G., {Kontar}, E.~P., \& {Sirenko}, O.~K. 2009, \apjl, 707, L45

\bibitem[{{Holman} {et~al.}(2011){Holman}, {Aschwanden}, {Aurass}, {Battaglia},
  {Grigis}, {Kontar}, {Liu}, {Saint-Hilaire}, \&
  {Zharkova}}]{2011SSRv..159..107H}
{Holman}, G.~D., {Aschwanden}, M.~J., {Aurass}, H., {et~al.} 2011, \ssr, 159,
  107

\bibitem[{{Kaplan} \& {Tsytovich}(1973)}]{1973plas.book.....K}
{Kaplan}, S.~A. \& {Tsytovich}, V.~N. 1973, {Plasma astrophysics}, ed. {Kaplan,
  S.~A.~\& Tsytovich, V.~N.}

\bibitem[{{Karlick{\'y}} \& {Ka{\v s}parov{\'a}}(2009)}]{2009A&A...506.1437K}
{Karlick{\'y}}, M. \& {Ka{\v s}parov{\'a}}, J. 2009, \aap, 506, 1437

\bibitem[{{Kontar}(2001{\natexlab{a}})}]{2001SoPh..202..131K}
{Kontar}, E.~P. 2001{\natexlab{a}}, Sol.~Phys., 202, 131

\bibitem[{{Kontar}(2001{\natexlab{b}})}]{2001A&A...375..629K}
{Kontar}, E.~P. 2001{\natexlab{b}}, \aap, 375, 629

\bibitem[{{Kontar}(2001{\natexlab{c}})}]{2001CoPhC.138..222K}
{Kontar}, E.~P. 2001{\natexlab{c}}, Computer Physics Communications, 138, 222

\bibitem[{{Kontar} {et~al.}(2011){Kontar}, {Brown}, {Emslie}, {Hajdas},
  {Holman}, {Hurford}, {Ka{\v s}parov{\'a}}, {Mallik}, {Massone}, {McConnell},
  {Piana}, {Prato}, {Schmahl}, \& {Suarez-Garcia}}]{2011SSRv..159..301K}
{Kontar}, E.~P., {Brown}, J.~C., {Emslie}, A.~G., {et~al.} 2011, \ssr, 159, 301

\bibitem[{{Kontar} \& {P{\'e}cseli}(2002)}]{2002PhRvE..65f6408K}
{Kontar}, E.~P. \& {P{\'e}cseli}, H.~L. 2002, \pre, 65, 066408

\bibitem[{{Krasovskii}(1978)}]{1978SvJPP...4R1267K}
{Krasovskii}, V.~L. 1978, Soviet Journal of Plasma Physics, 4, 1267

\bibitem[{{Krucker} {et~al.}(2011){Krucker}, {Hudson}, {Jeffrey}, {Battaglia},
  {Kontar}, {Benz}, {Csillaghy}, \& {Lin}}]{2011ApJ...739...96K}
{Krucker}, S., {Hudson}, H.~S., {Jeffrey}, N.~L.~S., {et~al.} 2011, \apj, 739,
  96

\bibitem[{{Li} {et~al.}(2006){Li}, {Robinson}, \&
  {Cairns}}]{2006PhPl...13i2902L}
{Li}, B., {Robinson}, P.~A., \& {Cairns}, I.~H. 2006, Physics of Plasmas, 13,
  092902

\bibitem[{{Lifshitz} \& {Pitaevskii}(1981)}]{1981phki.book.....L}
{Lifshitz}, E.~M. \& {Pitaevskii}, L.~P. 1981, {Physical kinetics}, ed.
  {Lifshitz, E.~M.~\& Pitaevskii, L.~P.}

\bibitem[{{Lin}(2006)}]{2006SSRv..124..233L}
{Lin}, R.~P. 2006, \ssr, 124, 233

\bibitem[{{Lin} {et~al.}(2003)}]{2003AdSpR..32.1001L}
{Lin}, R.~P. {et~al.} 2003, Advances in Space Research, 32, 1001

\bibitem[{{McClements}(1987)}]{1987A&A...175..255M}
{McClements}, K.~G. 1987, \aap, 175, 255

\bibitem[{{Mel'Nik} {et~al.}(1999){Mel'Nik}, {Lapshin}, \&
  {Kontar}}]{1999SoPh..184..353M}
{Mel'Nik}, V.~N., {Lapshin}, V., \& {Kontar}, E. 1999, \solphys, 184, 353

\bibitem[{{Melrose} \& {Cramer}(1989)}]{1989SoPh..123..343M}
{Melrose}, D.~B. \& {Cramer}, N.~F. 1989, \solphys, 123, 343

\bibitem[{{Muschietti} \& {Dum}(1991)}]{1991PhFlB...3.1968M}
{Muschietti}, L. \& {Dum}, C.~T. 1991, Physics of Fluids B, 3, 1968

\bibitem[{{Nishikawa} \& {Ryutov}(1976)}]{1976JPSJ...41.1757N}
{Nishikawa}, K. \& {Ryutov}, D.~D. 1976, Journal of the Physical Society of
  Japan, 41, 1757

\bibitem[{{Palastro} {et~al.}(2009){Palastro}, {Williams}, {Hinkel}, {Divol},
  \& {Strozzi}}]{2009PhPl...16i2304P}
{Palastro}, J.~P., {Williams}, E.~A., {Hinkel}, D.~E., {Divol}, L., \&
  {Strozzi}, D.~J. 2009, Physics of Plasmas, 16, 092304

\bibitem[{{Papadopoulos}(1975)}]{1975PhFl...18.1769P}
{Papadopoulos}, K. 1975, Physics of Fluids, 18, 1769

\bibitem[{{Papadopoulos} {et~al.}(1974){Papadopoulos}, {Goldstein}, \&
  {Smith}}]{1974ApJ...190..175P}
{Papadopoulos}, K., {Goldstein}, M.~L., \& {Smith}, R.~A. 1974, \apj, 190, 175

\bibitem[{{Priest} \& {Forbes}(2002)}]{2002A&ARv..10..313P}
{Priest}, E.~R. \& {Forbes}, T.~G. 2002, \aapr, 10, 313

\bibitem[{{Reid} {et~al.}(2011){Reid}, {Vilmer}, \&
  {Kontar}}]{2011A&A...529A..66R}
{Reid}, H.~A.~S., {Vilmer}, N., \& {Kontar}, E.~P. 2011, \aap, 529, A66+

\bibitem[{{Rowland} \& {Vlahos}(1985)}]{1985A&A...142..219R}
{Rowland}, H.~L. \& {Vlahos}, L. 1985, \aap, 142, 219

\bibitem[{{Ryutov}(1969)}]{1969JETP...30..131R}
{Ryutov}, D.~D. 1969, Soviet Journal of Experimental and Theoretical Physics,
  30, 131

\bibitem[{{Shibata}(1999)}]{1999Ap&SS.264..129S}
{Shibata}, K. 1999, \apss, 264, 129

\bibitem[{{Smith} \& {Sime}(1979)}]{1979ApJ...233..998S}
{Smith}, D.~F. \& {Sime}, D. 1979, \apj, 233, 998

\bibitem[{{Stepanov} \& {Tsap}(2002)}]{2002SoPh..211..135S}
{Stepanov}, A.~V. \& {Tsap}, Y.~T. 2002, \solphys, 211, 135

\bibitem[{{Sweet}(1958)}]{1958IAUS....6..123S}
{Sweet}, P.~A. 1958, in IAU Symposium, Vol.~6, Electromagnetic Phenomena in
  Cosmical Physics, ed. {B.~Lehnert}, 123--+

\bibitem[{{Syrovatskii} \& {Shmeleva}(1972)}]{1972AZh....49..334S}
{Syrovatskii}, S.~I. \& {Shmeleva}, O.~P. 1972, \azh, 49, 334

\bibitem[{{Tsiklauri}(2010)}]{2010SoPh..267..393T}
{Tsiklauri}, D. 2010, \solphys, 267, 393

\bibitem[{{Tsiklauri}(2011)}]{2011PhPl...18e2903T}
{Tsiklauri}, D. 2011, Physics of Plasmas, 18, 052903

\bibitem[{{Tsytovich}(1995)}]{1995lnlp.book.....T}
{Tsytovich}, V.~N. 1995, {Lectures on Non-linear Plasma Kinetics}, ed.
  {Tsytovich, V.~N.~\& ter Haar, D.}

\bibitem[{{Vedenov} {et~al.}(1967){Vedenov}, {Gordeev}, \&
  {Rudakov}}]{1967PlPh....9..719V}
{Vedenov}, A.~A., {Gordeev}, A.~V., \& {Rudakov}, L.~I. 1967, Plasma Physics,
  9, 719

\bibitem[{{Vedenov} \& {Velikhov}(1963)}]{1963JETP...16..682V}
{Vedenov}, A.~A. \& {Velikhov}, E.~P. 1963, Soviet Journal of Experimental and
  Theoretical Physics, 16, 682

\bibitem[{{Vlahos} \& {Papadopoulos}(1979)}]{1979ApJ...233..717V}
{Vlahos}, L. \& {Papadopoulos}, K. 1979, \apj, 233, 717

\bibitem[{{Yoon} {et~al.}(2005){Yoon}, {Rhee}, \& {Ryu}}]{2005PhRvL..95u5003Y}
{Yoon}, P.~H., {Rhee}, T., \& {Ryu}, C.-M. 2005, Physical Review Letters, 95,
  215003

\bibitem[{{Ziebell} {et~al.}(2001){Ziebell}, {Gaelzer}, \&
  {Yoon}}]{2001PhPl....8.3982Z}
{Ziebell}, L.~F., {Gaelzer}, R., \& {Yoon}, P.~H. 2001, Physics of Plasmas, 8,
  3982

\bibitem[{{Ziebell} {et~al.}(2011){Ziebell}, {Yoon}, {Pavan}, \&
  {Gaelzer}}]{2011ApJ...727...16Z}
{Ziebell}, L.~F., {Yoon}, P.~H., {Pavan}, J., \& {Gaelzer}, R. 2011, \apj, 727,
  16

\end{thebibliography}

\end{document}